\documentclass{ws-ijmpe}

\begin{document}

\title{Radio-loud Magnetars as Detectors for Axions and Axion-like Particles}

\author{\footnotesize Eduardo I. Guendelman}

\address{Physics Department, Ben-Gurion University, Beer-Sheva 84105, Israel; guendel@bgu.ac.il} 

\author{Doron Chelouche}

\address{Department of Physics, University of Haifa, Haifa 31905,  Israel ; doron@sci.haifa.ac.il} 

\maketitle

\begin{history}
\received{(received date)}
\revised{(revised date)}
\end{history}

\begin{abstract}

We show that, by studying the arrival times of radio pulses from highly-magnetized transient beamed sources, it may be possible to detect light pseudo-scalar particles, such as axions and axion-like particles, whose existence could have considerable implications for the strong-CP problem of QCD as well as the dark matter problem in cosmology. Specifically, such light bosons may be detected with a much greater  sensitivity, over a broad particle mass range, than is currently achievable by terrestrial experiments, and using indirect astrophysical considerations. The observable effect was discussed in Chelouche \& Guendelman (2009), and is akin to the Stern-Gerlach experiment: the splitting of a photon beam naturally arises when finite coupling exists between the electro-magnetic field and the axion field. The splitting angle of the light beams linearly depends on the photon wavelength, the size of the magnetized region, and the magnetic field gradient in the transverse direction to the propagation direction of the photons. If radio emission in radio-loud magnetars is beamed and originates in regions with strong magnetic field gradients, then splitting of individual pulses may be detectable. We quantify the effect for a simplified model for magnetars, and search for radio beam splitting in the 2\,GHz radio light curves of the radio loud magnetar XTE\,J1810-197. 

\end{abstract}

\section{Introduction}

The Peccei-Quinn mechanism (Peccei \& Quinn 1977) was devised to elegantly solve to the strong-CP problem of quantum chromodynamics (QCD). This was accomplished by postulating a new quantum field and a new class of particles associated with it. The particles are pseudo-scalars that couple very weakly to the electromagnetic (EM) field. It later became apparent that such particles, termed axions, could also provide a solution to the dark matter problem (Khlopov 1999 and references therein). Besides QCD axions there are also the putative axion-like particles (ALPs), which may be related to the quintessence field, and whose existence is predicted by many versions of string theory. To date, however, there is no evidence for the existence of such particles and it is not clear that the Peccei-Quinn solution actually works.

There is a longstanding interest in determining the physical properties of axions/ALPs. At present, laboratory experiments and astrophysical bounds imply that their  coupling constant to the electromagnetic field, $g<10^{-10}\,{\rm GeV}^{-1}$ (see Chelouche et al. 2009 for summary). Mass limits  are less stringent: if QCD axions are concerned, then their mass is probably $>10^{-6}$\,eV since otherwise the Universe would over-close, in contrast to observations. These limits, however, do not apply for ALPs. 

Here we follow the formalism given in Chelouche \& Guendelman (2009; see also Guendelman 2008a,b,c) who outlined a new effect that arises from the coupling between the electromagnetic field and the axion field. The effect has the advantage of having unique observational signatures, which can be visible down to very small values of the (unknown) coupling constant compared to those accessible by other methods. Below, we outline the effect of beam splitting and look for it in the radio light curve of a radio-loud magnetar.

\section{Splitting in in-homogenous magnetic fields}

The interaction term in the Lagrangian for the electromagnetic and the axion field is of the form 
\begin{equation}
\displaystyle L_{\rm int} = \frac{1}{4} g \tilde{F}^{\mu \nu}F_{\mu \nu}a =  g{\bf E}\cdot {\bf B} a
\end{equation}
where $E$ is the electric field (associated with the photon), $B$ the magnetic field, and $a$ the axion field. $g$ is the unknown coupling of particles to the EM field ($F_{\mu \nu}$, and its dual, $\tilde{F}^{\mu \nu}$). The full Lagrangian for the system can be written as
\begin{equation}
L= -\frac{1}{4}F^{\mu \nu}F_{\mu \nu} -\frac{1}{2}m_\gamma^2A^2 + \frac{1}{2} \partial_\mu a \partial ^\mu a -\frac{1}{2} m_a^2 a^2 +L_{\rm int}
\end{equation}
which is comprised of the free EM  Lagrangian (including an effective photon mass, $m_\gamma$, term which takes into account potential refractive index in the medium) and the Klein-Gordon equations for free particles having a rest-mass  $m_a$. In the absence of $L_{\rm int}$, photons and particles (e.g., axions/ALPs) are well defined energy states of the system. However, for finite coupling, the equation of motion for the photon-particle system takes the form (Raffelt \& Stodolsky 1988)\footnote{Unless otherwise stated, we work in natural units so that $\hbar=c=1$.},
\begin{equation}
\left [ {\bf k}^2 -\omega^2+\left \vert
\begin{array}{cc}
m_\gamma^2 & -gB_\| \omega \\
-gB_\| \omega & m_a^2
\end{array}
\right \vert \right ]  \left (
\begin{array}{l}
\gamma \\
a 
\end{array}
\right )=0,
\label{mat}
\end{equation}
where $\omega$ is the photon energy  and $B_\|$  the magnetic field in the direction of the photon polarization (the photon's $E$ field). Clearly, neither pure photon nor pure axion/ALP states are eigenstates of the system but rather some combination of them. 

\begin{figure}
\centerline{\psfig{file=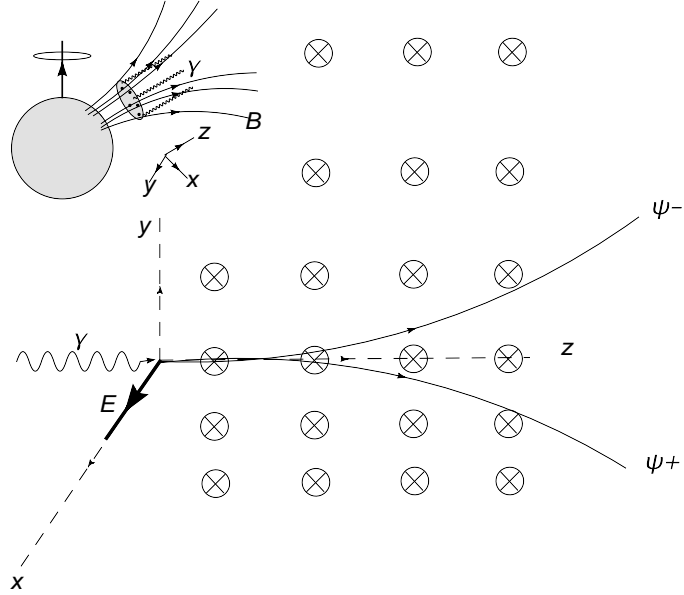,width=9cm}}
\caption{Photon-boson beams of different "charges"  (Eq. \ref{psi}) would be split along magnetic field gradients in a way similar to the Stern-Gerlach experiment. Photons propagate along the $z$-axis with their polarization along magnetic field lines ($x$-axis). A Schematic view of a magnetar is also shown. Magnetic field lines originate from the magnetic pole with plasma in its vicinity  emitting beamed radiation.  One possible orientation of the coordinate system is also shown.}
\label{1}
\end{figure}

Let us now focus on the limit 
\begin{equation}
\vert m_a^2-m_\gamma(\omega)^2 \vert \ll gB_\| \omega.
\label{cond2}
\end{equation}
This condition is met either near resonance where $m_\gamma^2 \simeq m_a^2$ or when both masses are individually smaller than $\sqrt{gB_\| \omega}$ (which limit is actually met is immaterial). The eigenstates of equation \ref{mat} are then given by 
\begin{equation}
\left \vert \psi \right >_- =  \left [ \left \vert  \gamma  \right > + \left \vert a \right > \right ]/\sqrt{2},~~\left \vert \psi \right >_+ =  \left [ \left \vert  \gamma  \right > - \left \vert a \right > \right ]/\sqrt{2}
\label{psi}
\end{equation}
where $\left \vert a \right >$ is the axion state and $\left \vert \gamma \right >$ is the photon state. The eigenvalues are $m_\pm^2= \pm gB_\| \omega$. By analogy with optics, these masses are related to effective refractive indices: $n_\pm=1+\delta n_\pm\simeq 1-{m_\pm^2}/{2\omega^2}$ (for $\vert \delta n_\pm\vert  \ll1$) meaning that different paths through a refractive medium would be taken by the rays. We note that there is no dependence on the particle or photon mass so long as equation \ref{cond2} is satisfied. 

In terms of the refractive index, and in complete analogy to mechanics, the equation of motion for a ray may be found by minimizing the action $\int d{\bf s} n({\bf s})$. It is straightforward to show (Chelouche \& Guendelman 2009) that the momentum imparted on each state is
\begin{equation}
\delta p_y^\pm = \mp  (g/2) \int dz  \left ( \partial B_x/\partial y  \right ),
\label{dpy}
\end{equation}
where $B_x=B_x(y,z)$, and is taken to be parallel to the photon polarization (Fig. 1). Clearly, each of the beams will be affected in a similar way while gaining opposite momenta so that the total momentum is zero and the classical wave packet travels in a straight line (along the $z$-axis). This effect is  analogous to the  Stern-Gerlach experiment (Fig. 1). In the limit $n_\pm \simeq 1$, the separation angle between the beams is
\begin{equation}
\delta \phi \simeq \frac{2\vert \delta p_y \vert}{p},
\label{theta}
\end{equation}
where $p$ is the beam momentum along the propagation direction, i.e., the $z$-axis. This expression holds for small splitting angles and assumes relativistic particles.  

\begin{figure}
\centerline{\psfig{file=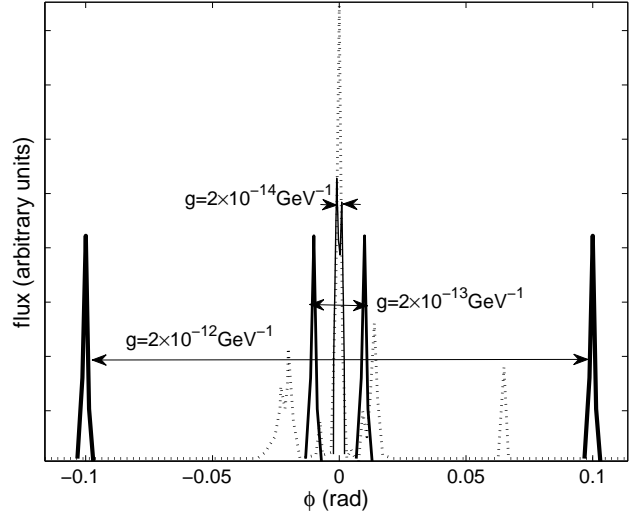,width=10cm}}
\caption{Splitting of the peak pulse at $\phi=0$\,rad in a single rotation light-curve of XTE\,J1810-197 (dotted line;  see Camilo et al. 2006) for several values of the coupling constant, $g$ and assuming $\lambda=1$\,m, $B_\|=10^{16}$\,G. Note the similar fluxes of the split signals whose sum corresponds to that of the original pulse. Looking for the effects of pulse-splitting in the radio light-curves of magnetars allows one to be considerably more sensitive to light bosons compared to terrestrial experiments. In particular, pulse splitting at meter wavelengths can be detected down to coupling constants $g \gtrsim 10^{-14}\,{\rm GeV}^{-1}$ for $m_a \ll 10^{-7}$\,eV (Chelouche \& Guendelman 2009). Observing at longer wavelengths (and assuming all other parameters are fixed) will proportionally increase the sensitivity to lower values of $g$.}
\label{splitting}
\end{figure}

The magnitude of splitting depends on the relative angle between the photon polarization and the magnetic field, as well as on the geometry (and strength) of the magnetic field, which are poorly understood in magnetars. To gain a qualitative understanding of the magnitude of the effect, we assume $B\sim \int dz  \left ( \partial B_x/\partial y \right ) $, where $B$ is the magnetic field in region through which the photon propagates. Taking current limits on the value of the coupling constant of $g\sim 10^{-10}\,{\rm GeV}^{-1}$, a photon frequency of 2\,GHz, and a typical magnetar magnetic field, $B\sim 10^{15}$\,G of we find that typical splitting phase between the photon beams is
\begin{equation}
\delta \phi \sim 1\left ( \frac{g}{10^{-10}\,{\rm GeV}^{-1}} \right ) \left ( \frac{B}{10^{15}\,{\rm G}} \right ) \left ( \frac{\omega}{2\,{\rm GHz}} \right )^{-1}\,{\rm rad}
\end{equation}
(here $\omega$ is the photon frequency in GHz, and $B$ is the magnetic field in Gauss). Provided that the intrinsic pulse emitted by the magnetar is narrower than the splitting angle, a double pulse is expected to appear due to the effect of splitting. In fact, in cases where the pulses are highly beamed, hence narrow in phase (see below), considerably smaller values of the coupling constant, $g$, may be probed. Furthermore, by choosing to work at lower frequencies, smaller coupling constants may be probed. This allows one to search for ALPs in a previously unexplored phase space using the radio light curves of radio-loud magnetars. Figure 2 shows an example for what a narrow radio pulse, typical of radio-loud magnetars (see below) would look like when split due to photon-particle coupling in a highly magnetized object ($B=10^{16}$\,G) observed at a radio frequency of 300MHz. Clearly, splitting may be discernible down to very low values of the coupling constant. 

We note that, depending on the strength of the magnetic field (and to some degree also on the poorly constrained plasma density), and the contribution of the vacuum birefringence term (Adler 1971) to the effective photon mass, two split pulses or one shifted pulse (with the shifting angle being $\delta \phi /2$) will be observed. In the latter case, light curves in two or more bands are required to detect the effect. The full treatment of such issues is beyond the scope of this contribution, and is discussed at some length in Chelouche \& Guendelman (2009). Below we consider only the effect of splitting when studying the (monochromatic) light curve of a radio-loud magnetar.

\section{The case of XTE\,J1810-197}

We adopt a pragmatic approach when searching for the effect of photon-axion coupling-induced splitting in magnetars. Given the large uncertainties in the physics of magnetars and their radio emission mechanisms, we do not know whether the effect of splitting or shifting should be observed. In addition, it is certainly possible that not all radio pulses originate from the same region in the magnetosphere, and while in some cases splitting will be observed, in other cases beam shifting will be the relevant effect. Similarly, if different pulses are emitted from different regions having different polarizations with respect to a complicated, tangled geometry of the field, then many different splitting or shifting angles are predicted. For these reasons, we aim to study the statistics of phase differences between pulses. Should all radio pulses be emitted from the same region, the data will reflect on the typical phase difference between pulses. This phase difference, which is induced by photon-particle coupling, may be discernible from other phase difference scales, which relate to the physics of the radio-emitting regions in the magnetar.

\begin{figure}
\centerline{\psfig{file=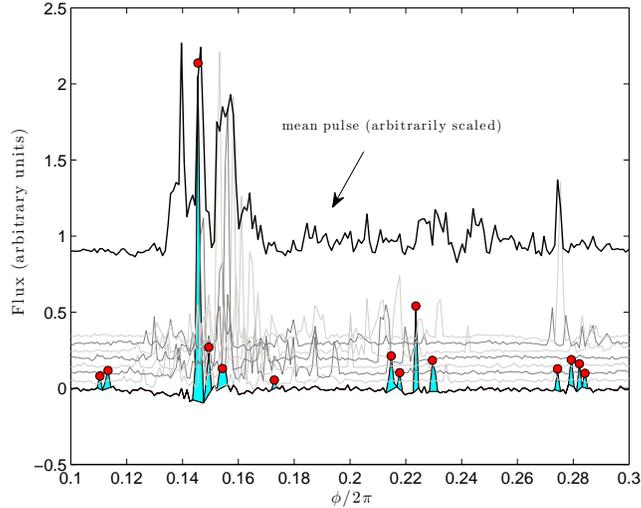,width=10cm}}
\caption{2\,GHz radio light curves (confined here for the orbital phases, $0.1<\phi/2\pi<0.3$) for individual revolutions of XTE\,J1810-197 are shown as dashed black curves, arbitrarily normalized for clarity (Camilo et al. 2006).  Significant peaks, as identified by our algorithm, are shown as cyan-shaded regions. The peak time is associated with the maximum of the pulse. In cases where several distinct pulses are observed (in a continuous, ridge-like form), individual maxima are recorded. The mean pulse (black thick curve) is also shown.}
\label{mag1}
\end{figure}

\begin{figure}
\centerline{\psfig{file=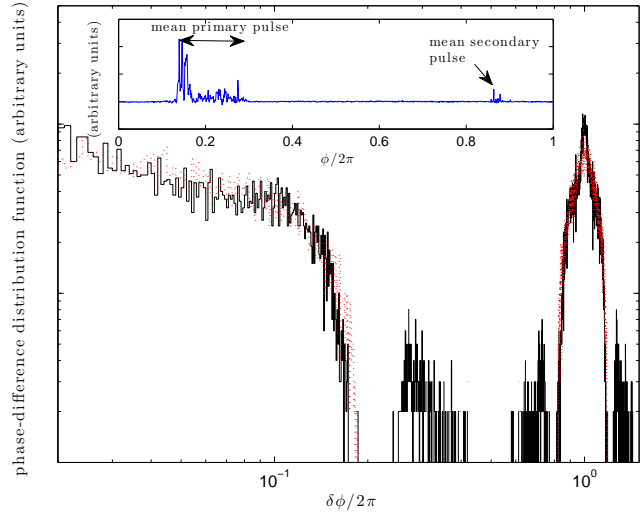,width=10cm}}
\caption{The peaks' phase difference distribution for XTE\,J1810-197 shows a peak at the orbital period ($\delta \phi/2\pi =1$) and smaller peaks corresponding to the separation between the primary and secondary peaks (see inset depicting the mean pulse over a stellar revolution). Below $\delta \phi/2\pi \sim 0.2$ (the width of the primary pulse), there is no clearly discernible timescale, and the distribution is qualitatively consistent with the radio emission peaks being drawn from a random process [red dashed line shows the results for a synthetic light curve model where random radio pulses were simulated for 40 object rotations within the primary pulse only (hence not showing primary-secondary peaks in the distribution function)].}
\label{mag2}
\end{figure}

We consider the 2\,GHz radio observations of the radio-loud magnetar XTE\,J1810-197 whose data were published by Camilo et al. (2006). A total of 40 object rotations were recorded, with the light curve of a few individual rotations shown in figure \ref{mag1}. As discussed in Camilo et al. (2006), the light curves are characterized by narrow transient radio pulses and, as such, are very different than those typically observed toward pulsars. When averaging the light curves of individual rotations, evidence for quasi-periodicity appears, whereby the bulk of the radio emission is confined to certain orbital phases, akin to the better studied pulsar phenomenon. 

Aiming to statistically study the difference in arrival phases of radio pulses, we first need to positively identify the numerous, potentially weak, narrow transient features in the light curves of XTE\,J1810-197. To this end, we devised the following "peak-finder" algorithm: for each light curve (rotation), we define the (initial) standard deviation of the light curve, $\sigma$. Only those peaks that satisfy $(f(t)-\left < f \right >)/\sigma>3$ [$f(t)$ is the time-series and $\left < f \right >$ is its mean], are identified as peaks, and are then removed from the observed light curve. A new standard deviation, $\sigma$, is calculated for the reduced light curve, and the peak identification algorithm is executed leading to new significant peaks being identified. The scheme iterates until $\sigma$ between successive iterations converges to better than 0.01\%. Figure \ref{mag1} shows the results of the peak identification algorithm for a light curve corresponding to a single stellar revolution. Clearly, all peaks lie well above the fluctuating background. The phase stamps of individual peaks are identified with their maxima. In cases were a multi-maxima ridge exists, the time stamps for individual maxima is recorded. We analyze the data from individual revolutions to be less sensitive to non-stationary effects in the light curve (e.g., a varying noise level between stellar revolutions).

All peaks from all stellar revolutions were identified and their phase stamps, relative to the first revolution, logged. We then evaluate the phase difference distribution taking into account all peak pairs. The results are shown in figure \ref{mag2}. A clear peak is observed, by definition, at around the stellar orbital period ($\delta \phi/2\pi=1$). Two small peaks at $\delta \phi/2\pi \sim 0.3,~0.7$ are due to the secondary pulse at $\phi/2\pi \sim 0.87$ (see the inset of Fig. \ref{mag2}). A second significant time-scale is apparent at a phase difference of $\delta \phi/2\pi \lesssim 0.2$. This scale roughly corresponds to the phase width of the mean main pulse (a second mean pulse exists at $\phi/2\pi \sim 0.87$ and is not shown here). Interestingly, we cannot positively identify any particular phase difference scale for $\delta \phi/2\pi <0.2$, as might be expected due to the effect of splitting. In fact, the distribution is qualitatively consistent with the predictions from a purely random origin for the radio pulses (see dotted line in Fig. \ref{mag2}). Further analysis is underway.

Based on our preliminary analysis, we cannot find supporting evidence for beam splitting in the 2GHz light curve of XTE\,J1810-197. There remains the open possibility that, for this object and this particular waveband, we are in the regime of beam shifting, and light curve comparison with {\it simultaneous} observations in other wavebands may be able to detect it. Given our limited understanding of magnetars and their radio emission processes, we do not claim to interpret our null result as a limit on the photon-ALP coupling constant, $g$, or on the existence of light bosons.

\section{Conclusions}

We show that the effect of beam splitting due to finite coupling between the axion field and the electromagnetic field (Chelouche \& Guendelman 2009) may be observable in the radio-light curves of radio-loud magnetars for a plausible range of values corresponding to the properties of magnetars and photon-to-axion coupling strength. The phase between the split pulses depends linearly on the magnetic field, the photon-particle coupling constant, and on the photon wavelength. As such, this effect can be used to detect axions and ALPs with much greater sensitivity than photon-axion/ALP oscillations. 

Our predictions indicate that, for narrow (beamed) radio pulses, the phase between the split pulses is likely to be $\lesssim 1$\,rad at 2\,GHz. Such a timescale will contribute to the statistics of phase differences between pulses, whose underlying form is determined by radio emission processes in the magnetar itself. 

Searching for discernible phase difference scales, which can be related to the beam-splitting effects, in the 2\,GHz light curve of XTE\,J1810-197, shows no clear characteristic phase scale in the range $0.1-1$\,rad. Interestingly, a preliminary analysis shows that the data is qualitatively consistent with narrow pulsed emission being drawn from a random process. While this could be used to shed light on the radio emission mechanism in magnetars, we cannot draw any conclusions at this stage concerning the existence of light bosons or their coupling to the electromagnetic field.

We thank Scott Ransom for providing us the 2\,GHz data for XTE\,J1810-197 in electronic form.

\end{document}